\documentclass[fleqn,usenatbib]{mnras}

\usepackage{newtxtext,newtxmath}

\usepackage[T1]{fontenc}

\DeclareRobustCommand{\VAN}[3]{#2}
\let\VANthebibliography\thebibliography
\def\thebibliography{\DeclareRobustCommand{\VAN}[3]{##3}\VANthebibliography}

\usepackage{graphicx}	
\usepackage{amsmath}	

\usepackage{natbib}
\RequirePackage{fix-cm}

\newcommand{\Msun}{M_{\odot}}

\newcommand{\cangle}{\overline{\theta}}  
\newcommand{\premorph}{\mathrm{(B/T)_{*,pre}}}  
\newcommand{\premorphsec}{\mathrm{(B/T)_{*,pre,sec}}}  
\newcommand{\postmorph}{\mathrm{(B/T)_{*,post}}}  
\newcommand{\fgas}{f_\mathrm{gas}}  
\newcommand{\fcoldgas}{f_\mathrm{cold\,gas}}  
\newcommand{\preMstars}{M_\mathrm{*,pre}}  
\newcommand{\signLext}{{\mathrm{sgn}}(L_\mathrm{ext})}  
\newcommand{\normLext}{\tilde{L}_\mathrm{ext}}  
\newcommand{\signLorb}{{\mathrm{sgn}}(L_\mathrm{orb})}  
\newcommand{\normLorb}{\tilde{L}_\mathrm{orb}}  

\title[How mergers shape morphology]{How mergers shape galaxy morphology in the IllustrisTNG simulation}

\author[Zeng et al.]{
Guangquan Zeng,$^{1,2}$
Lan Wang,$^{2,3}$\thanks{E-mail: wanglan@bao.ac.cn}
and Liang Gao $^{4,5}$
\\
$^{1}$Department of Physics, The Chinese University of Hong Kong, Sha Tin, N.T., Hong Kong, China\\
$^{2}$National Astronomical Observatories, Chinese Academy of Sciences, Beijing 100101, China \\
$^{3}$School of Astronomy and Space Science, University of Chinese Academy of Sciences, Beijing 100049, China\\
$^{4}$Institute for Frontiers in Astronomy and Astrophysics, Beijing Normal University, Beijing 102206, China \\
$^{5}$School of Physics and Microelectronics, Zhengzhou University, Zhengzhou 450001, China\\
}

\date{Accepted XXX. Received YYY; in original form ZZZ}

\pubyear{2015}

\begin{document}
\label{firstpage}
\pagerange{\pageref{firstpage}--\pageref{lastpage}}
\maketitle

\begin{abstract}
How galaxy mergers drive morphological evolution remains an open question.
Traditional views hold that major mergers produce elliptical galaxies, while minor mergers form dispersion-dominated components such as galactic bulges.
However, more recent work has challenged this simple picture, suggesting a more complex evolutionary scenario.
In this study, we use the IllustrisTNG cosmological simulation to investigate how mergers shape galaxy morphology, across a broad range of galaxy masses and merger mass ratios.
Our results show that the post-merger galaxy morphology is primarily determined by three factors: collision angle $\cangle$, cold gas fraction $\fcoldgas$, and pre-merger galaxy morphology $\premorph$.
Specifically, spiral-in mergers with large $\cangle$ increase the rotational support of the system, allowing the gas to settle into an extended disk.
When the system is rich in cold gas, star formation within the newly formed gas disk can further strengthen the disk-dominated structure of the remnant.
On the other hand, head-on mergers with small $\cangle$ typically disrupt ordered galactic motion, producing more dispersion-supported remnants.
Overall, we interpret our results within a unified picture of merger-driven morphological transformation in galaxies.

\end{abstract}




\section{Introduction}
\label{sec:Intro}

\defcitealias{2021MNRAS.507.3301Z}{Zeng+21}

Galaxy mergers have long been considered as a key mechanism that drives the morphological evolution of galaxies, particularly in transforming the disk structures into bulges or ellipticals \citep[][]{1972ApJ...178..623T, 1977egsp.conf..401T, 1978MNRAS.183..341W, 1979Natur.281..200F}.
In detail, major mergers between galaxies of comparable mass typically result in a dispersion-supported elliptical galaxy \citep[e.g.,][]{1982ApJ...259..103F, 1983MNRAS.205.1009N, 1988ApJ...331..699B, 1992ApJ...400..460H}.
On the other hand, minor mergers between galaxies of smaller mass-ratio would disturb the organized stellar motions, leading to the growth of bulge component \citep[e.g.,][]{2001A&A...367..428A, 2004A&A...418L..27B, 2005A&A...437...69B}.
This scenario has been implemented in many theoretical models to link mergers to the formation of bulges and elliptical galaxies \citep[e.g.,][]{1993MNRAS.264..201K, 2000MNRAS.319..168C, 2001MNRAS.320..504S, 2008MNRAS.391..481S, 2006MNRAS.365...11C, 2007MNRAS.375....2D, 2011MNRAS.413..101G, 2014MNRAS.439..264G, 2018MNRAS.481.1376Z}, successfully reproducing a range of observational results of galaxy morphology \citep[e.g.,][]{2011MNRAS.413..101G, 2018MNRAS.481.1376Z}.

However, hydrodynamical simulations reveal that morphological transformations during mergers are more complex than the traditional scenario suggests \citep[e.g.,][]{2017MNRAS.467.3083R, 2018MNRAS.478.3994C, 2018MNRAS.480.2266M, 2019MNRAS.485.2083W, 2019MNRAS.487.5416T, 2026A&A...706A.213R, 2026arXiv260721710N}.
Numerous studies have demonstrated that even major mergers can produce disk-dominated galaxies \citep[e.g.,][]{2005ApJ...622L...9S, 2006MNRAS.372..839N, 2006ApJ...645..986R, 2009MNRAS.398..312G, 2009ApJ...691.1168H, 2013MNRAS.430.1901H, 2016ApJ...821...90A, 2017MNRAS.470.3946S, 2020MNRAS.493.1375P, 2025ApJ...993L..28T, 2026arXiv260805835W}.
In gas-rich mergers, it has been shown that gas can reassemble into a rotational disk around the remnant, allowing the post-merger galaxy to evolve into a disk-dominated morphology \citep[e.g.,][]{2006ApJ...645..986R, 2007MNRAS.374.1479G, 2009MNRAS.398..312G, 2016ApJ...821...90A, 2017PASA...34...50F, 2020MNRAS.493.1375P, 2020MNRAS.494.5568J}.
Moreover, the bulge growth during mergers with sufficient gas could be significantly suppressed \citep[e.g.,][]{2009ApJ...691.1168H, 2011MNRAS.415.1051B, 2024MNRAS.528.2326S}.
Also, orbital configurations of mergers have been found to influence the morphology of remnant galaxies \citep[e.g.,][]{2003ApJ...597...21A, 2009ApJ...691.1168H, 2017MNRAS.467..179G, 2018MNRAS.479..141T, 2018MNRAS.480.2266M, 2018MNRAS.480L..18Z, 2022MNRAS.516.5404S, 2023MNRAS.523.3991Z, 2024arXiv240711444P}.
For example, prograde mergers are more likely to preserve the pre-existing disk structure of galaxies than retrograde mergers \citep[e.g.,][]{2022MNRAS.509.5062L, 2024RAA....24g5019H}.

In our previous study \citep[][hereafter \citetalias{2021MNRAS.507.3301Z}]{2021MNRAS.507.3301Z},
we examined the last major mergers of massive galaxies and found a clear correlation between morphological transformations and merger orbit types as represented by collision angle.
Specifically, head-on mergers typically produce dispersion-dominated remnants, whereas spiral-in mergers tend to preserve pre-existing disk structures and may even facilitate the formation of new rotation-supported disks.
This dependence on merger orbit type appears more significant than the dependence on other widely-discussed factors, such as gas fraction or prograde versus retrograde configurations.
However, the analysis in \citetalias{2021MNRAS.507.3301Z} was limited to major mergers in massive galaxies, leaving it unclear whether the strong correlation identified there also holds for minor mergers and for less massive galaxies.

Although various merger properties are found to influence galaxy morphological evolution, most studies, including \citetalias{2021MNRAS.507.3301Z}, have only assessed these effects qualitatively.
Therefore, it is still challenging to quantitatively determine which properties play a decisive role in different merger processes.
To address this, it would be beneficial to use the advanced statistical techniques to quantitatively assess how different merger properties contribute to the morphology of merger remnants.

Random forests provide an effective tool for such purposes.
As a widely used machine learning algorithm, random forests have gained popularity due to its strong performance in both classification and regression tasks \citep[][]{breiman2001random}.
In recent years, this method has been applied across many areas of astronomy, including galaxy morphology classification \citep[e.g.,][]{2018MNRAS.476.5516B, 2018MNRAS.474.5232S, 2021A&A...648A.122V}, estimation of galaxy mass and star formation rate \citep[e.g.,][]{2019A&A...622A.137B}, and identifying galaxy merger events \citep[e.g.,][]{2016MNRAS.458..226D, 2023MNRAS.519.4920G, 2024ApJ...976L...8R}.
For complex data sets, random forests can also quantify the importance of different features for predicting or explaining target properties \citep[e.g.,][]{2020MNRAS.492...96B, 2020MNRAS.499..230B, 2020ApJ...899...81C, 2023ApJ...942...54R, 2024ApJ...975...17J, 2024ApJ...975..234L, 2026MNRAS.548ag507C}.
For example, \citet{2024ApJ...975...17J} used random forests to quantitatively evaluate various observational galaxy properties and to determine those most responsible for quenched regions in nearby galaxies.

In this study, we use the random forest method to further investigate the qualitative findings of \citetalias{2021MNRAS.507.3301Z}, and quantify the relative importance of different merger properties.
These properties include cold gas fraction, prograde versus retrograde configurations, and merger orbit type, as well as the pre-merger morphologies of the galaxy pair, total gas fraction, external angular momentum, and orbital angular momentum.
We also extend our investigation across different galaxy mass ranges and merger mass ratios, aiming to identify which merger properties play dominant roles in morphological transformations.

This paper is organized as follows.
In Section \ref{sec:Methods}, we describe our methodology, including the TNG100-1 simulation, sample selection criteria, and random forest technique.
In Section \ref{sec:Quantify}, we quantify the relative importance of different merger properties on morphological transformations.
In Section \ref{sec:Mechanism}, we explore the underlying mechanisms by which the dominant merger properties shape galaxy morphology.
Our conclusions are summarized in Section \ref{sec:Summary}.

\section{Methodology}
\label{sec:Methods}

\subsection{TNG100-1 Simulation}
\label{subsec:Simulation}

The IllustrisTNG project \citep[][]{2018MNRAS.475..648P, 2018MNRAS.475..624N, 2018MNRAS.480.5113M, 2018MNRAS.475..676S, 2018MNRAS.477.1206N} is a suite of cosmological simulations designed to investigate the formation and evolution of galaxies.
These simulations implement the fiducial TNG galaxy formation model \citep[][]{2017MNRAS.465.3291W, 2018MNRAS.473.4077P}, and successfully reproduce a wide range of galaxy properties that are broadly consistent with observations \citep[e.g.,][]{2018MNRAS.479.4056W, 2018MNRAS.481.1950L, 2019MNRAS.483.4140R, 2019MNRAS.484.5587T}.
Of particular relevance to this study, the IllustrisTNG simulations also reproduce statistical trends in galaxy morphologies that agree well with observations \citep[e.g.,][]{2019MNRAS.487.5416T}.
Moreover, the evolution of galaxy sizes in IllustrisTNG closely follows the observational trends \citep[e.g.,][]{2018MNRAS.474.3976G}.

The IllustrisTNG project includes simulations with different box sizes, where larger volumes are run at lower resolution.
In this study, we use the TNG100-1 simulation, which provides a balance between resolution and statistical sample size.
Specifically, the TNG100-1 simulation evolves a periodic cubic volume with a side length of $110.7$ Mpc, initialized with $1820^3$ dark matter particles and an equal number of gas cells.
The mass resolution of dark matter particles is $7.5 \times 10^6 M_{\odot}$, and the baryonic mass resolution is approximately $1.4 \times 10^6 M_{\odot}$.
In IllustrisTNG simulations, non-star-forming gas is treated as a single-phase fluid with no cold phase.
When star formation sets in, the star-forming gas cell is modeled as a two-phase medium.
Moreover, the cold-phase mass fraction for simulated star-forming gas is generally greater than $90 \%$ \citep[see][]{2003MNRAS.339..289S, 2018MNRAS.473.4077P}.
Therefore, following previous works \citep[e.g.,][]{2018ApJS..238...33D, 2021MNRAS.507.4445N, 2021MNRAS.507.3301Z, 2024MNRAS.532.2558Z}, the cold gas content of a galaxy is defined in this study as gas with a non-zero star formation rate.

In IllustrisTNG simulations, dark matter halos and subhalos are identified in each snapshot using the \texttt{Friends-of-Friends} (\texttt{FoF}) algorithm \citep[][]{1985ApJ...292..371D} and the \texttt{Subfind} algorithm \citep[][]{2001MNRAS.328..726S, 2009MNRAS.399..497D}.
Subhalos containing non-zero stellar mass are classified as galaxies.
The merger histories (i.e., the so-called merger trees) of dark matter subhalos and their associated galaxies are constructed using the \texttt{Sublink} algorithm \citep[][]{2015MNRAS.449...49R}, which identifies the progenitors and the unique descendant of each subhalo across simulation snapshots.
At each snapshot, the main progenitor of a given subhalo or galaxy is defined as the most massive progenitor.
By following the progenitors across successive snapshots, one can reconstruct the main progenitor branch and the assembly history of a subhalo or galaxy.

Based on the subhalo/galaxy merger trees constructed in TNG100-1, a merger is considered to occur when two galaxies share a common descendant.
Previous studies have shown that, in hydrodynamical simulations, as two galaxies approach one another, the less massive galaxy can lose a significant fraction of its stellar mass due to tidal stripping before coalescence\footnote{In this paper, the term \textit{coalescence} refers specifically to the simulation snapshot at which two merging galaxies are identified as a single one for the first time. In contrast, the term \textit{merger} refers to the entire event, including both the pre-coalescence interaction and subsequent relaxation during the time window considered.} \citep[e.g.,][]{2019MNRAS.485.2083W, 2020MNRAS.493.1375P}.
Therefore, to obtain the merger mass ratio that is less affected by the interaction, in this study, for each merging galaxy, we trace its main progenitor branch and adopt the maximum stellar mass along this branch to calculate the ratio \citep[see discussions in][]{2015MNRAS.449...49R}.

To trace galaxy morphological changes before and after mergers,
following previous studies \citep[e.g.,][]{2018MNRAS.480.2266M, 2025A&A...699A.374W, 2026RAA....26k5009P}, we adopt a time window of $\pm 1$ Gyr relative to coalescence, which roughly captures the full merger process\footnote{We note that this approach may not be universally applicable to all merger types \citep[e.g.,][]{2008ApJ...675.1095J, 2010MNRAS.404..575L, 2025ApJ...986..201X}, therefore we check this for our samples and confirm that this time window choice does not affect our main conclusions in this work.}.
Also, in following analysis, the subscripts $_\texttt{pre}$ and $_\texttt{post}$ are used to denote the pre-merger and post-merger properties measured at $-1$ Gyr and $+1$ Gyr relative to coalescence, respectively.
The subscript $_\texttt{sec}$ refers to the less massive secondary galaxy in each merger pair, and a property written without a subscript refers to either the more massive primary galaxy or the merger event as a whole.

\subsection{Sample Selection of Galaxy Mergers}
\label{subsec:Samples}

To investigate the morphological transformations induced by mergers across a broad range of mass ratios and galaxy masses, we first extract the merger histories of all galaxies at $z=0$ in TNG100-1 and record the stellar mass ratio of each merger. 
We then select our sample using the following criteria:
\begin{itemize}
\item To ensure sufficient resolution of both the merger event and the resulting morphological change, we include only mergers with a stellar mass ratio of $\mu_* > 1:10$, and with a pre-merger primary galaxy stellar mass of $\preMstars > 10^9 \Msun$, which corresponds to roughly 1,000 stellar particles in the TNG100-1 simulation.
\item To focus on the effect of individual mergers, we exclude any event accompanied by another merger with $\mu_* > 1:10$ within the $\pm 1$ Gyr window, thereby minimizing the contamination from other nearly-simultaneous mergers.
\end{itemize}

\begin{figure}
\centering
\includegraphics[width=1.0\columnwidth]{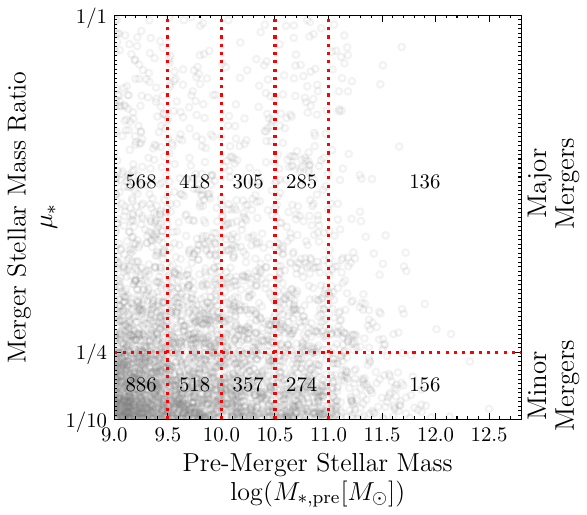}
\caption{
Mergers selected from the TNG100-1 simulation.
For each merger represented by a gray circle, the horizontal axis shows the pre-merger stellar mass of the primary galaxy, and the vertical axis shows the merger stellar mass ratio.
Based on various values of these two parameters, as indicated by the red dashed lines, the total merger sample is further divided into ten sub-samples for detailed analysis.
Specifically, the vertical red lines separate five mass bins with $\log (\preMstars [\Msun])$ in the ranges of [$ 9.0, 9.5$], [$9.5, 10.0$], [$10.0, 10.5$], [$10.5, 11.0$], and $> 11.0$.
The horizontal red line separates major mergers with $\mu_* > 1:4$ from minor mergers with $1:10 < \mu_* < 1:4$.
The number of mergers in each sub-sample is indicated by the black annotations.
}
\label{fig:tng100-1_mergers_subsample_division}
\end{figure}

After applying these selection criteria, we obtain a final sample of $3,903$ mergers.
The distributions of pre-merger stellar masses and merger ratios for the whole sample are shown in Fig.~\ref{fig:tng100-1_mergers_subsample_division}.

We further classify major mergers with $\mu_* > 1:4$ and minor mergers with $1:10 < \mu_* < 1:4$, to distinguish their different roles in driving morphological evolution.
In addition, since galaxy properties vary significantly with stellar mass, we divide the mergers into five intervals according to the primary galaxy pre-merger stellar mass, with $\log (\preMstars [\Msun])$ in the range of [$9.0, 9.5$], [$9.5, 10.0$], [$10.0, 10.5$], [$10.5, 11.0$], and $> 11.0$.
This yields ten sub-samples in total, covering both major and minor mergers across different mass bins, 
as indicated in Fig.~\ref{fig:tng100-1_mergers_subsample_division}.

\subsection{Merger Properties for Random Forest Regression}
\label{subsec:RF}

\begin{table*}
\centering
\caption{
Input and target features for the random forest regression used in this work.
The input features consist of various merger properties, and the target feature is the post-merger morphology (see text for detailed definitions).
After training, the random forest regression yields the relative importance of each input feature in predicting the target, enabling a direct comparison of how individual merger properties affect the post-merger morphology.
}
\label{tab:merger_features}
\footnotesize 
\setlength{\tabcolsep}{15pt} 
\renewcommand{\arraystretch}{1.5} 
\begin{tabular}{ll}
\hline
\hline
\textbf{Target Feature} & \textbf{Description} \\
\hline
$\postmorph$ & Post-merger morphology \\ 
\hline
\hline
\textbf{Input Features} & \textbf{Description} \\
\hline
$\cangle$ & Collision angle of the merger, averaged over the $1$ Gyr window before coalescence \\
$\premorph$ & Pre-merger morphology of the primary (more massive) galaxy in the merger pair \\
$\premorphsec$ & Pre-merger morphology of the secondary (less massive) galaxy in the merger pair \\
$\fgas$ & Total gas fraction of the merger \\
$\fcoldgas$ & Cold gas fraction of the merger \\
$\signLext$ & Direction of the external angular momentum (either $+1$ for prograde, or $-1$ for retrograde) \\
$\normLext$   & Magnitude of external angular momentum, normalized by the total mass of primary galaxy \\
$\signLorb$ & Direction of the orbital angular momentum (either $+1$ or $-1$)  \\
$\normLorb$ & Magnitude of the orbital angular momentum, normalized by the total mass of primary galaxy \\
\hline
\hline
\end{tabular}
\end{table*}

As introduced in Section \ref{sec:Intro}, this study uses random forest regression to quantify the relative importance of various merger properties in driving galaxy morphological transformation.
Specifically, we select the post-merger morphology as the target feature, while a set of merger properties serve as the input features for training the random forest regressions.

The target feature and the merger properties considered as input features are listed below and summarized in Table~\ref{tab:merger_features}.

\begin{itemize}

\item \textbf{Target Feature}
\begin{itemize}
\item \textbf{Post-merger galaxy morphology}:
We measure the bulge-to-total stellar mass ratio $\mathrm{(B/T)_{*}}$ of the remnant galaxy at $1$ Gyr after coalescence, to represent the final post-merger morphology.
This serves as the regression target for the random forest.

Following previous studies \citep[e.g.,][]{2009MNRAS.396..696S, 2012MNRAS.423.1726S}, we quantify galaxy morphology based on the kinematic circularity parameter $\epsilon$, which is the ratio of a simulation particle's specific angular momentum to that of a circular orbit at the same radius.
For a given galaxy, the bulge-to-total mass ratio $\mathrm{B/T}$ is then defined as twice the mass fraction of the counter-rotating component with $\epsilon < 0$.
This ratio ranges from $0$ to $1$, corresponding to galaxies that are purely rotation-supported and dispersion-supported, respectively.
\end{itemize}

\item \textbf{Input Features}
\begin{itemize}
\item \textbf{Collision angle}:
For each snapshot within $1$ Gyr before coalescence, we calculate the acute angle between the position and velocity vectors of the secondary galaxy relative to the primary galaxy.
We then take the average over all snapshots as the collision angle, to characterize the merger orbit type:
\begin{equation}
    \cangle = \frac{1}{n} \sum_{i=1}^{n} \theta_{i},
\end{equation}
where $\theta_{i}$ represents the angle at each snapshot before coalescence.
Under this definition, which follows our previous work \citep[\citetalias{2021MNRAS.507.3301Z},][]{2025A&A...699A.374W, 2026RAA....26k5009P}, a large collision angle $\cangle$ close to 90$^{\circ}$ indicates a spiral-in merger orbit, whereas a small $\cangle$ close to 0$^{\circ}$ indicates that the merger orbit is head-on.

\item \textbf{Pre-merger galaxy morphology}:
We adopt the bulge-to-total stellar mass ratio $\mathrm{(B/T)_{*}}$ measured at $1$ Gyr before coalescence as the pre-merger morphology for both the primary and secondary galaxies, represented by $\premorph$ and $\premorphsec$ respectively.

\item \textbf{Gas fraction}:
The gas fraction of the merger system is defined as the ratio of gas mass to the sum of gas mass and stellar mass.
We calculate the total gas fraction at the snapshot $1$ Gyr before coalescence as:
\begin{equation}
    \fgas = \frac{\mathrm{total \; gas \; mass}}{\mathrm{total \; gas \; mass} + \mathrm{total \; stellar \; mass}},
\end{equation}
and the cold gas fraction as:
\begin{equation}
\fcoldgas = \frac{\mathrm{total \; cold \; gas \; mass}}{\mathrm{total \; cold \; gas \; mass} + \mathrm{total \; stellar \; mass}},
\end{equation}
As introduced in Section \ref{subsec:Simulation}, we adopt cold gas mass as the sum of the star-forming gas cells.
These simulated star-forming gas cells typically consist of more than 90\% cold-phase gas, as modeled by the IllustrisTNG sub-grid physics (see \citealp{2018MNRAS.473.4077P} and also \citealp{2003MNRAS.339..289S} for details).

\item \textbf{Orbital configuration}:
To further characterize the merger orbital configuration, we measure the external angular momentum and the orbital angular momentum at $1$ Gyr before coalescence.
Following previous studies \citep[e.g.,][]{2018MNRAS.480.2266M}, the external angular momentum is calculated as:
\begin{equation}
    L_\mathrm{ext} = |L_\mathrm{sec}|\cos{({\theta}_{L_\mathrm{pri},L_\mathrm{sec}})} + |L_\mathrm{orb}|\cos{({\theta}_{L_\mathrm{pri},L_\mathrm{orb}})},
	\label{eq:L_ext}
\end{equation}
where $L_\mathrm{pri}$ and $L_\mathrm{sec}$ are the stellar angular momentum vector of the primary and secondary galaxies, respectively.
The orbital angular momentum $L_\mathrm{orb} = \text{M}_\text{sec}(\textbf{r} \times \textbf{v})$ is calculated by the merger orbit of secondary galaxy with respect to the primary.
Here, $\theta_{L_\mathrm{pri},L_\mathrm{sec}}$ is the angle between the angular momentum vectors of the primary and secondary galaxies,
and $\theta_{L_\mathrm{pri},L_\mathrm{orb}}$ is the angle between the primary galaxy angular momentum and the orbital angular momentum.
A prograde merger is defined as merger with a positive external angular momentum, whereas a retrograde merger corresponds to a negative value.

In this study, we adopt the sign of the external angular momentum $\signLext$ (i.e., $+1$ for prograde, $-1$ for retrograde), and its normalized magnitude $\normLext = | L_\mathrm{ext} | / M_\mathrm{pri}$ as input features.
Similarly, the direction and normalized magnitude of the orbital angular momentum, i.e., $\signLorb$ and $\normLorb = | L_\mathrm{orb} | / M_\mathrm{pri}$, are also included as input features.
\end{itemize}

\end{itemize}

We adopt a random forest regression comprising 1,000 decision trees to perform the fitting for a given merger sub-sample, then derive the relative importance of each input feature.
These feature importance values are normalized to sum to $1$, allowing for a quantitative comparison of the contribution of each input feature to the target.
Specifically, a higher feature importance indicates a more significant influence of that property on the post-merger galaxy morphology.
In contrast, a feature importance lower than $1 / N_{\text{features}}$ (i.e., roughly $0.1$ in this study) suggests that the property has negligible influence on the post-merger galaxy morphology.

A brief overview of the random forest algorithm and the calculation of feature importance is provided in Appendix~\ref{sec:Appendix_A}.

\section{Dominant Merger Properties in Reshaping Morphology}
\label{sec:Quantify}

In this section, we use the random forest regression to quantify the relative importance of various merger properties in driving galaxy morphological transformation during mergers.
We begin with the major mergers involving massive galaxies, then extend the analysis to different types of mergers and across various galaxy masses.
This allows us to determine the dominant factors driving morphological transformation in different mergers.


\subsection{Last Major Merger of Massive Galaxies}
\label{subsec:QuantifyZeng2021}

\begin{figure}
\centering
\includegraphics[width=0.9\columnwidth]{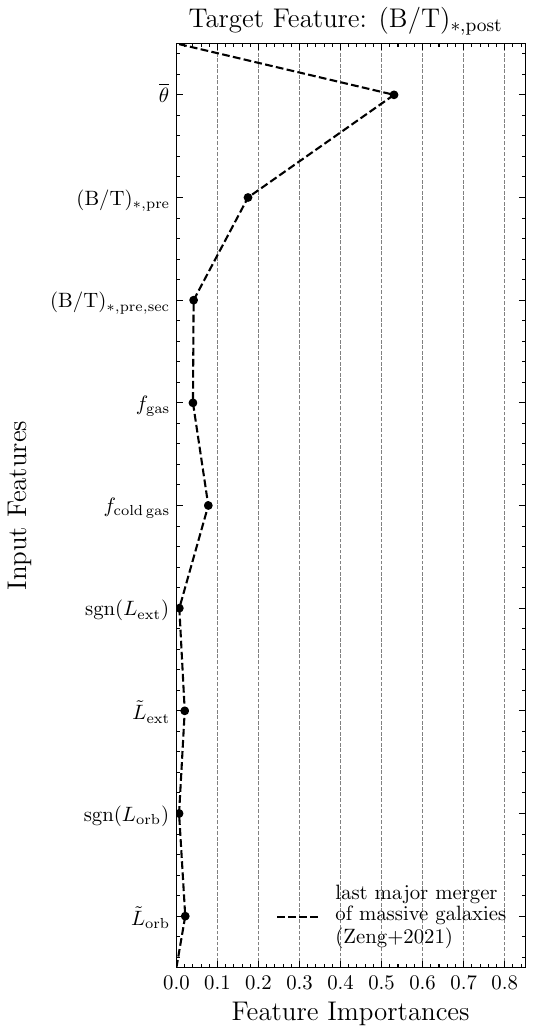}
\caption{
Feature importances of various merger properties obtained from the random forest regression, for the last major merger events at $z<1$ of massive galaxies.
The sum of all feature importances is normalized to $1$, with a higher feature importance indicating a more significant influence on the post-merger galaxy morphology.
For these mergers, the collision angle $\cangle$ and the pre-merger morphology $\premorph$ primarily determine the post-merger morphology, in agreement with the qualitative findings of \citetalias{2021MNRAS.507.3301Z} (see their Figure 7).
}
\label{fig:tng100-1_Zeng2021_mergers_compare_quant}
\end{figure}

In \citetalias{2021MNRAS.507.3301Z} we found that, for the last major merger of present-day massive galaxies, the morphological change correlates more strongly with the merger orbit type represented by collision angle, rather than with the cold gas fraction or the prograde versus retrograde configuration.
Here, we use random forest regression to carry out a quantitative analysis, in which we measure the relative importance of these properties as well as other merger properties that may affect morphological transformation.

Based on the galaxy merger sample constructed in Section \ref{subsec:Samples}, for massive galaxies with $M_{*} > 8 \times 10^{10} M_{\odot}$ at $z=0$, we pick out their last major merger at $z<1$, to be consistent with \citetalias{2021MNRAS.507.3301Z}.
For these selected major mergers, we train the random forest regression to evaluate the relative importance of all input features (as listed in Table~\ref{tab:merger_features}) in predicting the post-merger morphology.
The resulting feature importances are presented in Fig.~\ref{fig:tng100-1_Zeng2021_mergers_compare_quant}.

As shown in Fig.~\ref{fig:tng100-1_Zeng2021_mergers_compare_quant}, the collision angle $\cangle$ is the most important factor in driving morphological transformation during these major mergers, with a feature importance exceeding $0.5$.
Quantitatively this confirms the dominant role of collision angle in reshaping galaxy morphology as reported in \citetalias{2021MNRAS.507.3301Z}.
Beyond that, the pre-merger morphology $\premorph$ of the primary galaxy also has a notable importance of nearly $0.2$.
This is naturally expected,  since the initial morphology sets the starting point for the transformation.
Besides the above two properties, all others have feature importances below $0.1$ or even close to $0$, suggesting that their individual contributions to the overall morphological transformation are negligible.

Results in Fig.~\ref{fig:tng100-1_Zeng2021_mergers_compare_quant} quantitatively confirm the findings of \citetalias{2021MNRAS.507.3301Z} that the collision angle, together with pre-merger morphology, largely determines the galaxy morphology after major merger (see their Fig. 7 for comparison).
This consistency demonstrates the random forest technique can effectively and quantitatively compare how different merger properties impact the galaxy morphological transformation.
Therefore, in the following subsections we extend this analysis to major mergers in various galaxy masses, and also to minor mergers.

\subsection{Major Mergers across Different Stellar Mass Ranges}
\label{subsec:QuantifyMajor}

\begin{figure*}
\hspace{-0.4cm}
\resizebox{13.5cm}{!}{\includegraphics{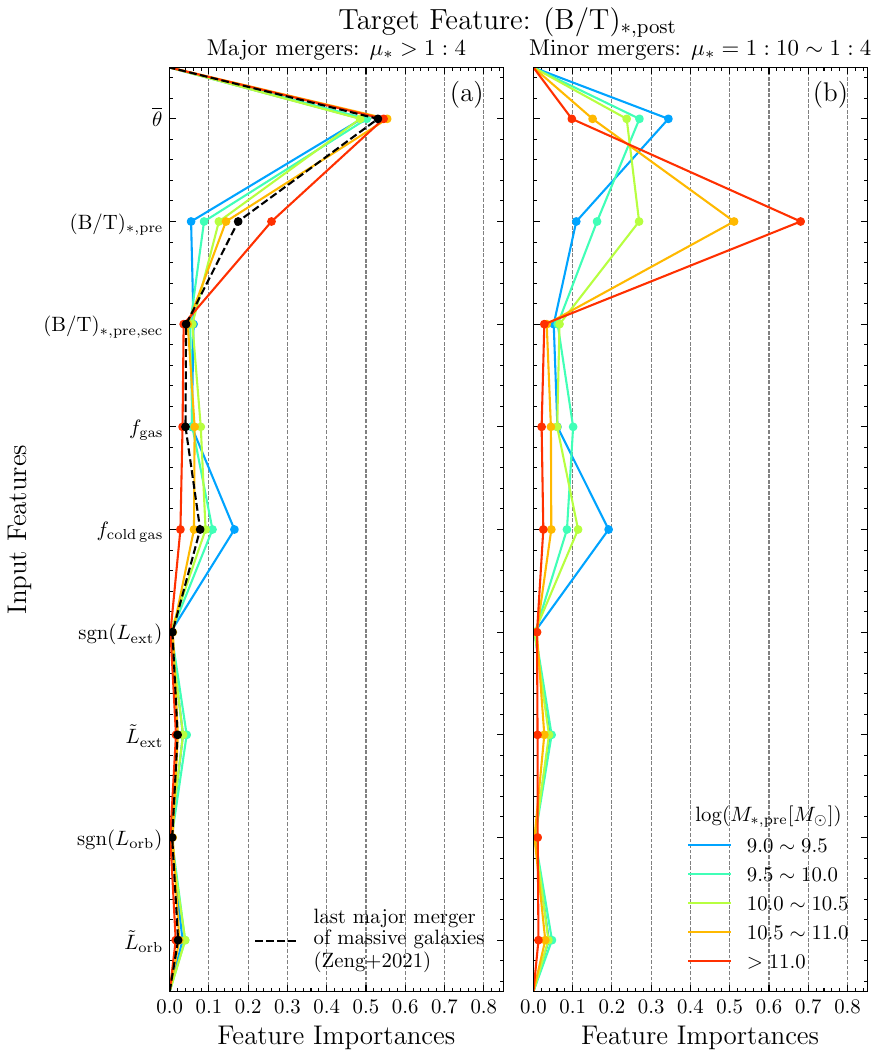}}
\caption{
Feature importance of various merger properties in predicting the post-merger morphology for (a) major mergers, and (b) minor mergers.
Similar to Fig.~\ref{fig:tng100-1_Zeng2021_mergers_compare_quant}.
In each panel, different colored lines represent the results for different pre-merger stellar mass bins.
For reference, the black dashed line in panel (a) gives the result for the last major mergers of massive galaxies, identical to that shown in Fig.~\ref{fig:tng100-1_Zeng2021_mergers_compare_quant}.
For both major and minor mergers across all stellar mass ranges, only three properties have a considerable influence on the post-merger morphology: collision angle $\cangle$, pre-merger morphology $\premorph$, and cold gas fraction $\fcoldgas$.
Together, these three properties primarily determine the post-merger galaxy morphology, with the relative importance varying across different merger ratios and stellar masses.
}
\label{fig:tng100-1_redshift_0d0_to_20d0_different_mergers_compare_quant}
\end{figure*}

In this subsection, we apply the random forest method to five sub-samples of major mergers with different stellar masses, selected as described in Section \ref{subsec:Samples} and shown in Fig.~\ref{fig:tng100-1_mergers_subsample_division}.
For each of the sub-samples, we train the random forest regression to predict the post-merger morphology, and to identify the key properties governing the morphological transformation.

The resulting feature importances are presented in the panel (a) of Fig.~\ref{fig:tng100-1_redshift_0d0_to_20d0_different_mergers_compare_quant}.
The colored solid lines, ranging from blue to red, correspond to sub-samples with increasing pre-merger stellar masses.

Comparing feature importances across sub-samples, a clear result is that for all major mergers, regardless of the pre-merger galaxy mass, the collision angle $\cangle$ is always the most important factor determining the post-merger morphology.
The feature importances of $\cangle$ exceed $0.5$ in all mass bins, which is significantly higher than that of any other property.
Moreover, the feature importance of $\cangle$ in major mergers shows no obvious variation across mass bins.

Beyond $\cangle$, panel (a) shows that only two other properties can reach a feature importance above $0.1$ in some mass bins: the pre-merger galaxy morphology $\premorph$, and the cold gas fraction $\fcoldgas$. 
Their relative importance varies across mass bins.
In the most massive sub-sample, the second most important factor after $\cangle$ is $\premorph$.
In less massive bins, however, the importance of $\premorph$ gradually declines, while that of $\fcoldgas$ increases.
In the least massive sub-sample, $\fcoldgas$ becomes the second most important property.

Besides $\cangle$, $\premorph$ and $\fcoldgas$, panel (a) in Fig.~\ref{fig:tng100-1_redshift_0d0_to_20d0_different_mergers_compare_quant} shows that all remaining merger properties have a feature importance consistently below $0.1$, and in some cases close to $0$, indicating their negligible impact on the post-merger morphology.

These quantitative results demonstrate that, for major mergers, the collision angle $\cangle$ is always the primary determinant of the post-merger morphology across all mass ranges.
In addition, both the cold gas fraction $\fcoldgas$ and the pre-merger morphology $\premorph$ show considerable influence, although their relative importance varies with galaxy mass systematically.
Overall, for major mergers, only these three properties substantially affect the post-merger morphology.

\subsection{Minor Mergers across Different Stellar Mass Ranges}
\label{subsec:QuantifyMinor}

Following the analysis above, we now apply the random forest method to minor mergers with $1:10 < \mu_* < 1:4$ across different galaxy masses.
The resulting feature importances are shown in panel (b) of Fig.~\ref{fig:tng100-1_redshift_0d0_to_20d0_different_mergers_compare_quant}.

Panel (b) shows that, unlike in major mergers, the collision angle is not always the dominant feature in determining galaxy post-merger morphology for minor mergers.
Moreover, the relative importance of different properties varies far more with stellar mass in minor mergers than in major mergers.
For galaxies more massive than $10^{10.5} \, M_{\odot}$, $\premorph$ primarily determines the post-merger morphology, with an importance reaching $0.5$ or even $0.7$, while $\cangle$ plays a weak and sub-dominant role.

As galaxy mass decreases, the importance of $\premorph$ declines, while that of $\cangle$ increases.
In the lowest mass bin, $\cangle$ becomes the primary factor affecting post-merger morphology.
Meanwhile, the importance of $\fcoldgas$ also rises with decreasing galaxy mass, and becomes the second most important feature in the lowest mass bin, surpassing the effect of $\premorph$.

Overall, as in major mergers, the dominant properties determining the post-merger morphology in minor mergers remain the same three:
collision angle $\cangle$, pre-merger galaxy morphology $\premorph$, and cold gas fraction $\fcoldgas$.
Although their relative importance varies with merger mass ratios and galaxy masses, these three properties markedly determine the post-merger galaxy morphology.

\section{How Merger Properties Reshape Morphology}
\label{sec:Mechanism}

Having identified the key features that determine the final post-merger morphology, we now investigate in detail how the three dominant properties, i.e., collision angle $\cangle$, pre-merger morphology $\premorph$, and cold gas fraction $\fcoldgas$, operate and interact to jointly drive the morphological transformation.

In the following, we look at the dependence of morphological changes on these three properties in different sub-samples, to study why their relative importances vary across different merger mass ratios and galaxy masses.
In Section \ref{subsec:CaseStudy}, we conduct case study to further understand the physical processes driving morphological transformation
In Section \ref{subsec:UnifiedPicture}, we combine the results into a unified scenario for how merger properties reshape galaxy morphology in different cases.

\subsection{Reshaping in Major Mergers}
\label{subsec:ReshapingMajor}

\begin{figure*}
\hspace{-0.4cm}
\resizebox{17.8cm}{!}{\includegraphics{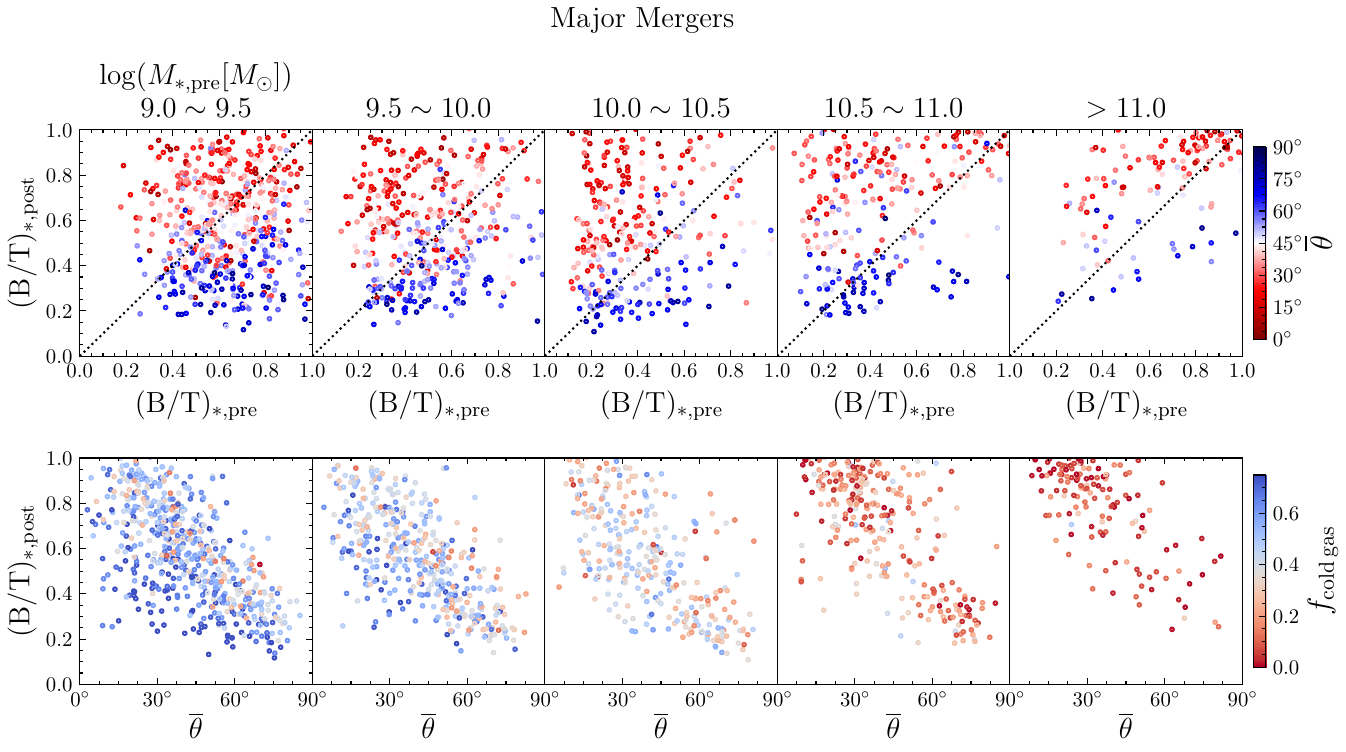}}
\caption{
Dependence of galaxy post-merger morphology on the three key merger properties, $\premorph$, $\cangle$, and $\fcoldgas$, in major mergers of five different mass bins.
\textit{Upper panels:} The horizontal and vertical axes represent pre-merger morphology $\premorph$ versus post-merger morphology $\postmorph$, with symbols color-coded by the collision angle $\cangle$.
Bluer colors indicate more spiral-in mergers, while redder colors represent more head-on mergers.
\textit{Bottom panels:} The horizontal and vertical axes show $\cangle$ versus $\postmorph$, with symbols color-coded by the cold gas fraction $\fcoldgas$.
Bluer colors indicate more cold gas-rich mergers, whereas redder colors indicate more cold gas-poor mergers.
This figure shows that for major mergers, $\cangle$ is the primary determinant of the post-merger morphology, with spiral-in orbits producing disk galaxies, while head-on orbits yield ellipticals.
The sub-dominant features are $\fcoldgas$, which further promotes disk formation in low-mass galaxy mergers where it is abundant, and $\premorph$, which helps preserve pre-existing structure in massive galaxies.
}
\label{fig:tng100-1_redshift_0d0_to_20d0_different_major_mergers_morph_change}
\end{figure*}

In the upper panels of Fig.~\ref{fig:tng100-1_redshift_0d0_to_20d0_different_major_mergers_morph_change}, we examine the dependence of the post-merger morphology on both $\premorph$ and $\cangle$ for major mergers in different mass bins. 
Symbols are color-coded by the collision angle $\cangle$, with bluer colors indicating more spiral-in mergers and redder colors indicating  more head-on mergers.
As shown, in all mass bins, spiral-in major mergers with large $\cangle$ tend to lie in the lower part of each panel, producing disk-like galaxies, whereas head-on mergers with small $\cangle$ mostly occupy the upper region producing elliptical galaxies.
This clear separation is consistent with the quantitative results in Fig.~\ref{fig:tng100-1_redshift_0d0_to_20d0_different_mergers_compare_quant}, showing that for major mergers, the collision angle $\cangle$ is the primary determinant of the post-merger morphology across all mass ranges.

Across the upper panels of Fig.~\ref{fig:tng100-1_redshift_0d0_to_20d0_different_major_mergers_morph_change}, from left to right, the data points become more concentrated along the diagonal dashed line, indicating a stronger correlation between pre- and post-merger morphologies.
This trend reflects the growing influence of $\premorph$ on the post-merger morphology with increasing stellar mass, consistent with the rising feature importance of $\premorph$ shown in panel (a) of Fig.~\ref{fig:tng100-1_redshift_0d0_to_20d0_different_mergers_compare_quant}.
For galaxies less massive than $10^{10} M_{\odot}$, the post-merger morphology is almost unaffected by their pre-merger morphology.
This suggests that more massive galaxies are more resistant to structural disruption and reshaping.
But in general, the morphological transformation of major mergers is largely determined by $\cangle$.

In the lower panels of Fig.~\ref{fig:tng100-1_redshift_0d0_to_20d0_different_major_mergers_morph_change}, we further check the dependence of the post-merger morphology on $\fcoldgas$.
Here the symbols are color-coded by $\fcoldgas$, with bluer and redder colors indicating cold gas-rich and -poor mergers, respectively.
In all mass bins, it is seen again that there exists a strong correlation between the post-merger morphology and $\cangle$, with larger $\cangle$ (more spiral-in orbit) corresponding to smaller $\postmorph$ (more disky remnant), and the correlation is clear across all mass bins.

Another noticeable trend is that, the symbols become redder from left to right in the lower panels,
which means major mergers of less massive galaxies typically involve more cold gas, whereas that of more massive galaxies involve much less.
Also, it is seen that the cold gas fraction spans a wider range in less massive galaxies.
In the leftmost lower panel, for a given $\cangle$, mergers with higher $\fcoldgas$ tend to produce a more disk-dominated remnant.
This indicates that, the presence of abundant cold gas would further promote the formation of a disk-like post-merger galaxy, consistent with earlier studies \citep[e.g.,][]{2005ApJ...622L...9S, 2006MNRAS.372..839N, 2006ApJ...645..986R}.

For more massive galaxies, where mergers involve progressively less cold gas, the influence of $\fcoldgas$ becomes weaker.
When the galaxy mass exceeds about $10^{10.5} \, M_{\odot}$, nearly all major mergers in our sample have $\fcoldgas < 0.2$.
This is likely due to the efficient feedback from supermassive black holes in massive galaxies in the IllustrisTNG simulation \citep[see e.g.,][]{2017MNRAS.465.3291W}.
For these mergers, the post-merger morphology is largely independent of the precise cold gas fraction, and is instead determined primarily by the collision angle $\cangle$ and, to a lesser extent, the pre-merger morphology.

\subsection{Reshaping in Minor Mergers}
\label{subsec:ReshapingMinor}

\begin{figure*}
\hspace{-0.4cm}
\resizebox{17.8cm}{!}{\includegraphics{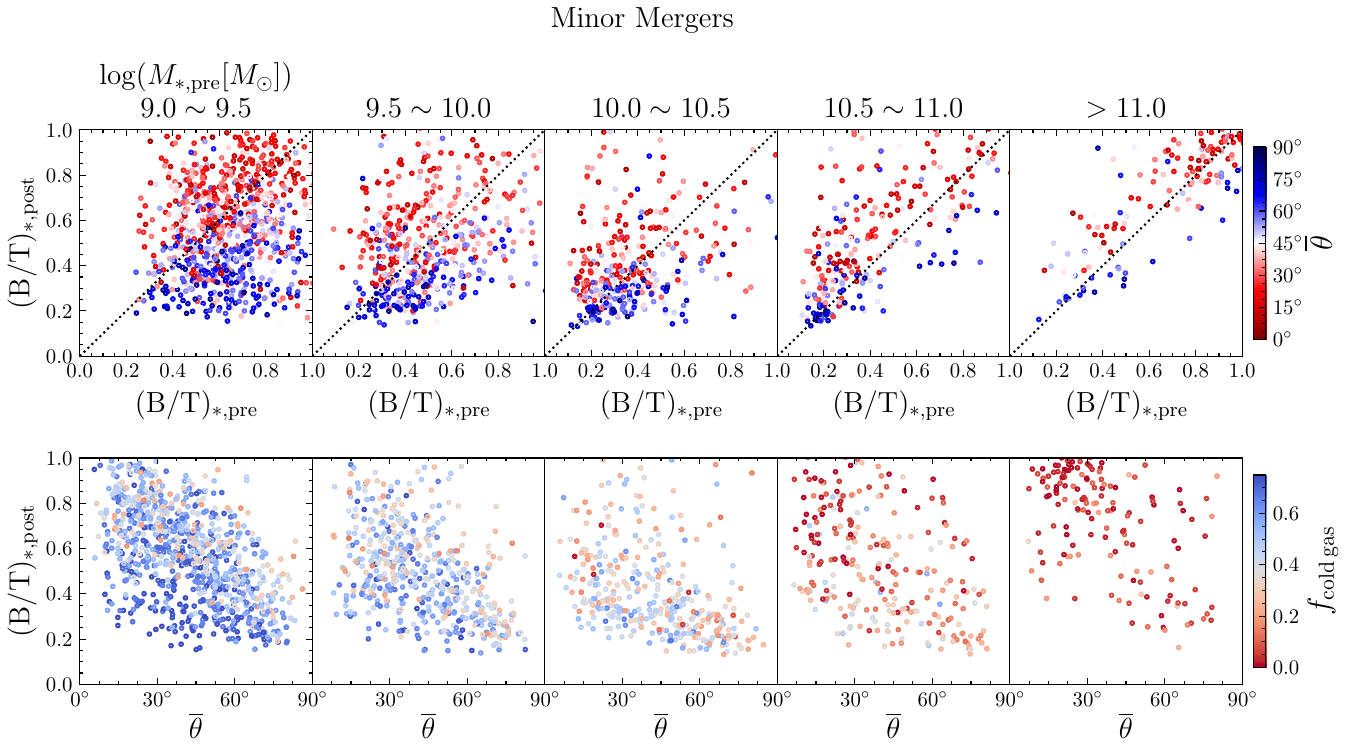}}
\caption{
Similar to Fig.~\ref{fig:tng100-1_redshift_0d0_to_20d0_different_major_mergers_morph_change}, but for minor mergers in different mass bins.
For minor mergers, the general trends for dependence of the post-merger morphology on $\cangle$, $\premorph$ and $\fcoldgas$ are similar to those for major mergers, except that the dependence is stronger on $\premorph$ and weaker on $\cangle$.
}
\label{fig:tng100-1_redshift_0d0_to_20d0_different_minor_mergers_morph_change}
\end{figure*}

In this subsection, we analyze minor mergers across different galaxy mass bins, following the same approach as for major mergers in subsection~\ref{subsec:ReshapingMajor}, and show the results in Fig.~\ref{fig:tng100-1_redshift_0d0_to_20d0_different_minor_mergers_morph_change}.

Compared with the major merger results in Fig.~\ref{fig:tng100-1_redshift_0d0_to_20d0_different_major_mergers_morph_change}, the upper panels of Fig.~\ref{fig:tng100-1_redshift_0d0_to_20d0_different_minor_mergers_morph_change} show a weaker dependence of post-merger morphology on $\cangle$, especially for more massive galaxies.
Nevertheless, the same trend holds, where spiral-in minor mergers tend to produce more disky galaxies and head-on minor mergers tend to form more elliptical galaxies.
By contrast, the dependence of post-merger morphology on $\premorph$ is much stronger in minor mergers, particularly for galaxies more massive than $10^{9.5} M_{\odot}$.
This is expected, as the smaller infalling galaxy in a minor merger is less capable of perturbing and thus reshaping the host structure.

As for the dependence on cold gas fraction, the lower panels of Fig.~\ref{fig:tng100-1_redshift_0d0_to_20d0_different_minor_mergers_morph_change} show trends also similar to those seen for major mergers in Fig.~\ref{fig:tng100-1_redshift_0d0_to_20d0_different_major_mergers_morph_change}.
In low-mass galaxies, where $\fcoldgas$ spans a wider range, cold gas-rich minor mergers tend to produce a more disky remnant for a given $\cangle$.
On the other hand, minor mergers in massive galaxies involve relatively low $\fcoldgas$, resulting in a relatively small amount of new star formation during the merger.
Combined with the fact that a small infalling galaxy has limited ability to redistribute pre-existing stars, the morphology of the remnant in minor mergers of massive galaxies is thus dominated by the original morphology.

\subsection{Case Study: A Spiral-in Cold Gas-Rich Major Merger}
\label{subsec:CaseStudy}


\begin{figure*}
\hspace{-0.4cm}
\resizebox{16.8cm}{!}{\includegraphics{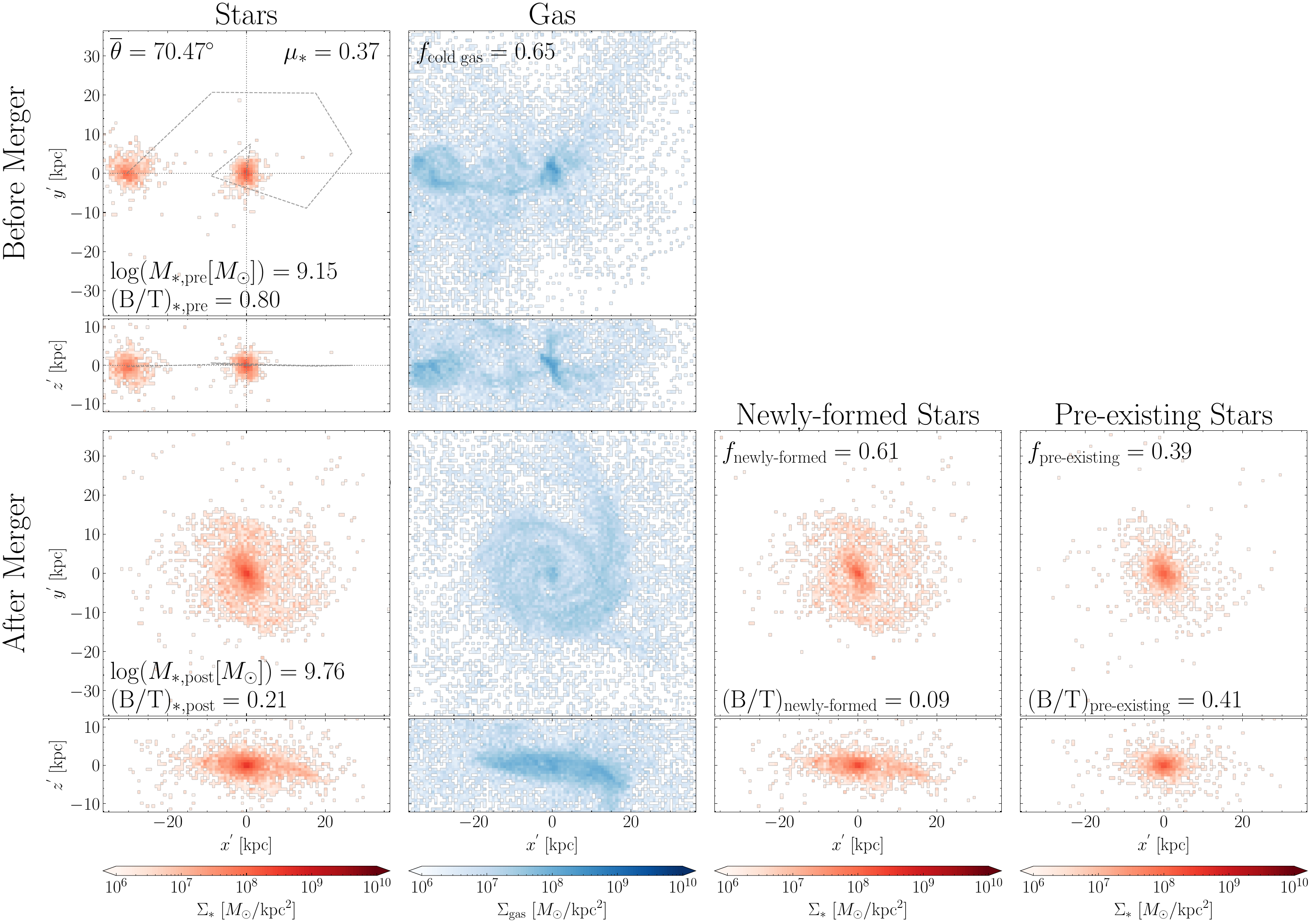}}
\caption{
A typical case of a cold gas-rich ($\fcoldgas = 0.65$) major merger ($\mu_* = 0.37$) with a spiral-in orbit ($\cangle = 70.47^{\circ}$), illustrating how cold gas and collision angle affect the galaxy morphological transformation.
The upper rows show the density projections of stars (red) and gas (blue) before the merger, viewed from two directions:  the $x' - y'$ and $x' - z'$ projection planes correspond to viewing angles perpendicular and parallel to the merger orbital plane, respectively.
The merger orbit of this case is indicated by the gray dashed lines, and various properties before merger are also labeled.
In bottom rows, the left panels show the density projections of stars and gas after the merger.
In addition, the right panels show the density projections of the stars newly formed during the merger and of the pre-existing stars, with their mass fractions and $\mathrm{B/T}$ values labeled.
}
\label{fig:tng100-1_case_study_cold_gas_rich_spiral_in_major_merger}
\end{figure*}

Why does the collision angle play a dominant role in reshaping galaxy morphologies during mergers?
How does the cold gas affect the final morphology of merger remnants in detail?
To gain more insight into how these properties determine the post-merger galaxy morphology, we visually inspected the evolutionary processes of a large number of our merger samples.

In Fig.~\ref{fig:tng100-1_case_study_cold_gas_rich_spiral_in_major_merger}, we present a typical case of a cold gas-rich major merger with a spiral-in orbit, illustrating how cold gas and collision angle affect the galaxy morphological transformation.
As shown in the upper panels, before the merger, the primary galaxy has a dispersion-dominated elliptical morphology with $\premorph = 0.80$.
However, after the spiral-in merger with $\cangle = 70.47^{\circ}$, the galaxy evolves into a disk galaxy with $\postmorph = 0.21$, as shown in the bottom left panels of Fig.~\ref{fig:tng100-1_case_study_cold_gas_rich_spiral_in_major_merger}.
Moreover, the disk plane of the post-merger galaxy aligns clearly with the merger orbital plane.
As also shown in the gas projection panels, after undergoing such a spiral-in merger, the gas content of this merging galaxy pair is arranged along the orbital plane, forming a notable gas disk.

In the bottom right panels, the density projections of stars formed during the merger and pre-existing stars are furtner shown separately.
In this case, the newly-formed stars during merger constitute the majority of the total stellar mass in the post-merger galaxy.
Moreover, these stars exhibit a prominent disk structure with $\mathrm{B/T}$ as low as $0.09$, dominating the post-merger morphology.
Meanwhile, the pre-existing stars of both the primary and secondary galaxies also tend to become more disk-like during this spiral-in merger, ending up as part of the post-merger remnant, with a final $\mathrm{B/T}$ value of $0.41$.

The case shown in Fig.~\ref{fig:tng100-1_case_study_cold_gas_rich_spiral_in_major_merger} demonstrates that, for a spiral-in merger, the stars and gas in the merging galaxy pair are rearranged along the orbital plane, forming rotationally-supported disk structures in that plane.
Given sufficient cold gas, an extended cold gas disk forms and drives intense star formation within it, further making the post-merger galaxy considerably disk-like.

Applying the same analysis to other individual mergers in our sample, we find that they follow the similar evolutionary pattern.
In general, a spiral-in merger orbit typically redistributes stars and gas into the orbital plane, forming more rotationally-supported structures than previously existed.
When abundant cold gas is available, stars would form within the established cold gas disk, thus producing a large new stellar disk component that makes the galaxy even more disk-like.
When cold gas is limited, the redistribution of pre-existing stars also makes the galaxy more disk-like, but generally to a lesser extent.

In contrast, when a head-on merger occurs, the ordered motion (if present) of the stars and gas in the galaxies can be significantly disrupted, driving the post-merger galaxy toward a more dispersion-dominated structure, especially when cold gas is limited.
However, in cold gas-rich mergers, the abundant cold gas can play an additional role.
In some cases, even in a head-on merger, a large cold gas fraction can allow a gas disk to settle and form stars, making the remnant more disk-dominated than its cold-gas-poor counterpart.
This explains why, in low-mass galaxies where mergers are generally cold gas-rich, a fraction of head-on mergers still produce disk-like remnants, as shown in Fig.~\ref{fig:tng100-1_redshift_0d0_to_20d0_different_major_mergers_morph_change} and Fig.~\ref{fig:tng100-1_redshift_0d0_to_20d0_different_minor_mergers_morph_change}.

While head-on collisions align more closely with the traditional view that galaxy mergers perturb or even destroy galactic disks, the spiral-in orbits offer an alternative pathway that allows disk survival and subsequent growth, or even disk regrowth.
In summary, collision angle $\cangle$, cold gas fraction $\fcoldgas$, and also pre-merger morphology $\premorph$, jointly determine the final morphology of a post-merger galaxy.
The detailed dependence of post-merger morphology on these three properties, across major and minor mergers and different galaxy mass bins (as shown in Section \ref{sec:Quantify} and Sections \ref{subsec:ReshapingMajor} and \ref{subsec:ReshapingMinor}), can be naturally explained as arising from the varying values of these properties across different cases, which lead to varying contributions to the final morphological outcomes.

\subsection{A Unified Picture}
\label{subsec:UnifiedPicture}


The analysis above points to a unified picture of how mergers reshape galaxy morphology.

Fundamentally, the galaxy merger is a process of redistributing the galactic material, including both stars and gas, in both spatial and kinematic ways.
Each merger property affects this redistribution distinctly, and together they determine the post-merger morphology.

\begin{itemize}
\item \textbf{Collision angle $\cangle$} describes the mode of the merger-induced material redistribution.
In spiral-in mergers, stars and gas are guided onto the orbital plane, reassembling the system into a rotational disk-like configuration.
In head-on mergers, by contrast, ordered motion in stars and gas would be more easily disrupted or even completely destroyed, making the system more dispersion-dominated.

\item \textbf{Pre-merger morphology $\premorph$} represents the resistance to the material redistribution.
When the redistribution is weak, as in minor mergers, and/or little star formation occurs during the merger, as in cold gas-poor mergers, the pre-existing stars would dominate the post-merger outcome.
The resulting morphology therefore more closely resembles the original one.

\item \textbf{Cold gas fraction $\fcoldgas$} determines how many stars are born during the merger, and facilitates disk formation by regulating the number of newly-formed stars that settle into a disk afterward.
In a cold gas-rich spiral-in merger, cold gas is channeled into a disk configuration.
Meanwhile the abundant star formation within this cold gas disk can dominate the morphological transformation, overwhelmingly producing a disk-dominated post-merger galaxy.
Even in head-on mergers with abundant cold gas, a gaseous disk can still settle and form stars, making the remnant more disk-like than in cold gas-poor cases.
\end{itemize}

Overall, these properties jointly determine the post-merger galaxy morphology,
and their different combinations give rise to different evolutionary paths in the merger-induced transformation.

\section{Conclusions and Discussions}
\label{sec:Summary}

In this work, using the TNG100-1 simulation, we investigate quantitatively which merger properties dominate and how they drive the merger-induced galaxy morphological transformation, across a wide range of galaxy masses and merger mass ratios.

We find that the post-merger morphology is primarily determined by only three factors: collision angle $\cangle$, cold gas fraction $\fcoldgas$, and pre-merger morphology $\premorph$, although their relative importance varies with stellar mass and mass ratio (Fig. \ref{fig:tng100-1_redshift_0d0_to_20d0_different_mergers_compare_quant}).
Furthermore, how these factors jointly shape the post-merger morphology (Figs. \ref{fig:tng100-1_redshift_0d0_to_20d0_different_major_mergers_morph_change} and \ref{fig:tng100-1_redshift_0d0_to_20d0_different_minor_mergers_morph_change}) can be interpreted within a unified physical picture:

Fundamentally, galaxy mergers redistribute the galactic material, including both stars and gas, in both a spatial and a kinematic manner.

(1) For spiral-in mergers, the orbit guides the material to re-orient and settle along the orbital plane, leading both stars and gas into a rotation-supported disk-like configuration.
Moreover, if the merging pair is rich in cold gas, the whole system not only forms a gaseous disk during the merger, but also triggers star formation within this newly established gas disk, further reinforcing a post-merger disk-dominated morphology (see Fig. \ref{fig:tng100-1_case_study_cold_gas_rich_spiral_in_major_merger} for an example).

(2) Conversely, head-on mergers typically disrupt the pre-existing ordered motion in stars and gas, driving the galaxy toward a more dispersion-dominated morphology.
Besides, cold gas introduces an additional effect that cold gas-rich systems tend to form more disk-dominated remnants.
In some cases where head-on mergers involve abundant cold gas, a gaseous disk can still settle and form stars, making the remnant more disk-like than in cold gas-poor cases.

This scenario applies to both major and minor mergers, except that the merger-induced rearrangement depends naturally on the mass ratio and is generally more significant for more equal-mass mergers.
Due to their smaller mass ratios, minor mergers are considerably less violent than major mergers, and as a result, their post-merger morphology remains more closely tied to the pre-merger state.

Our results highlight the key role of collision angle in shaping the morphological outcome of galaxy mergers.
Recently, \citet{2026RAA....26k5009P} compared the merger orbits in TNG simulations with those in corresponding dark-matter-only simulations, and found that their systematic differences in collision angle are generally small, indicating that the collision angle is largely not affected by the presence of baryons.
Also using the TNG simulation, \citet{2025A&A...699A.374W} performed a detailed galaxy decomposition into bulge, disk, warm component, and hot inner stellar halo, and also found the collision angle strongly affects how these components respond to mergers.
Furthermore, Gan et al. (in preparation) have also confirmed the dominant effect of collision angle in the EAGLE simulation \citep[][]{2015MNRAS.446..521S, 2015MNRAS.450.1937C}, which adopts different subgrid physics from the TNG simulations.
Taken together, the dominant role of the collision angle in driving merger-induced morphological transformations appears robust.

We note that the collision angle is distinct from other merger orbital parameters that are more commonly discussed in the literature, such as the orbital angular momentum.
Our upcoming paper will further explore the underlying physics that makes collision angle unique in determining the post-merger galaxy morphology.

We also note that this study focuses only on the overall trends in how different merger properties affect the morphological evolution.
We have not attempted, and it is unlikely we are able, to accurately predict the post-merger galaxy properties from any given pre-merger conditions.
As shown in Figs. \ref{fig:tng100-1_redshift_0d0_to_20d0_different_major_mergers_morph_change} and \ref{fig:tng100-1_redshift_0d0_to_20d0_different_minor_mergers_morph_change}, a few cold gas-rich spiral-in mergers produce elliptical-like remnants, whereas cold gas-poor head-on mergers can also yield disk galaxies.
These outliers may point to physical factors not accounted for in this study, which could be explored in future work.

\section*{Acknowledgements}


We acknowledge the support from the National Natural Science Foundation of China (Grant No.12588202), the National SKA Program of China (No.2022SKA0110201), the National Key Research and Development Program of China (No.2023YFB3002501), and the Strategic Priority Research Program of Chinese Academy of Sciences, (grant No. XDB0500203).
\section*{Data Availability}


The IllustrisTNG simulations are publicly available and accessible at \url{https://www.tng-project.org/data/}.
The data produced in this work will be shared upon reasonable request to the corresponding author.



\bibliographystyle{mnras}
\bibliography{references}




\appendix

\section{Random Forest Regression}
\label{sec:Appendix_A}

Random forest is a widely used machine learning method \citep[][]{breiman2001random} that improves the task performance for both classification and regression by ensembling multiple individual decision trees \citep[][]{kotsiantis2013decision}.
Technically, random forest mitigates the common overfitting issue of single decision tree by leveraging bagging sampling and random feature selection.
First, the bagging sampling generates multiple subsets from the training data via bootstrap sampling (i.e., sampling with replacement), and uses these multiple subsets to build multiple decision trees.
Then by aggregating the predictions from all decision trees (e.g., majority voting for classification, or averaging for regression), the overfitting can be effectively reduced.
Second, random feature selection considers only a subset of features at each split node of each decision tree, which further increases randomness among trees and avoids improper dependence on specific features.
Through the random selection of both training samples and features, random forest prevents over-reliance on any particular set of samples or features, making it less prone to overfitting and more robust to outliers.

For a given set of samples, each described by various properties, random forest regression can be used to identify which properties predominantly determine a target property.
Specifically, one property is selected as the target feature, and the remaining properties serve as input features for the regression model.
After training on the provided data, the random forest regressor quantifies the importance of each input feature in predicting the target.
These feature importance scores are normalized to sum to $1$, allowing a quantitative comparison of the contribution of each feature to the target.

This study follows the common practice of calculating feature importance based on node impurity reduction within decision trees.
For a decision tree in a regression task, the node impurity can be measured using the mean-square error (MSE):
\begin{equation}
\text{MSE} = \frac{1}{N} \sum_{i=1}^{N} (y_i - \bar{y})^2,
\end{equation}
where $N$ is the number of samples at the current node, $y_i$ is the actual value of the $i$-th sample, and $\bar{y}$ is the mean of the samples.

For each decision tree, the reduction in node impurity achieved by splitting on feature $j$ can be calculated as:
\begin{equation}
\Delta I_j = I_{\text{before}} - \left( \frac{N_{\text{left}}}{N} I_{\text{left}} + \frac{N_{\text{right}}}{N} I_{\text{right}} \right),
\end{equation}
where $I_{\text{before}}$ is the impurity before the split, $I_{\text{left}}$ and $I_{\text{right}}$ are the impurities of the left and right child nodes after the split, and $N_{\text{left}}$ and $N_{\text{right}}$ are the corresponding numbers of samples in these child nodes.

Finally, the feature importance for feature $j$ can be measured by its cumulative contribution across all decision trees:
\begin{equation}
I_j = C \sum_{i=1}^{m} \Delta I_{j,i},
\end{equation}
where $m$ is the total number of decision trees in the random forest, $\Delta I_{j,i}$ is the impurity reduction attributed to feature $j$ in the $i$-th tree, and $C$ is a normalization constant ensuring that the sum of all feature importances equals $1$.

In this study, we use random forest regression to quantify the influence of different merger properties on the post-merger galaxy morphology.
The input features of the random forest correspond to the various merger properties, while the target feature is the post-merger galaxy morphology.
Each decision tree in the random forest is trained on a subset of the data obtained via bootstrap sampling, with also only a random subset of the input features considered at each split.
 Once trained, the random forest would provide feature importances for the input features, which quantify the contribution of each to predicting the target.
By comparing these feature importances, we can thus quantitatively compare the influence of various merger properties on the post-merger galaxy morphology.





\bsp	
\label{lastpage}
\end{document}